\documentclass[
 preprint,
preprint
preprintnumbers,
 amsmath,amssymb,
 aps,
prd,
showkeys
]{revtex4-1}

\onecolumngrid

\usepackage{graphicx}
\usepackage{bm}

\usepackage{braket}
\usepackage{slashed}
\usepackage{extarrows}
\usepackage{enumerate}
\usepackage{orcidlink}

\newcommand{\tr}{\text{tr}}
\newcommand{\Eq}[1]{Eq.~(\ref{#1})}

\begin{document}

\preprint{ADP-26-12/T1309}

\title{On the Origins of the Strong CP Problem}

\author{Anthony G. Williams}
\affiliation{%
  ARC Centre of Excellence for Dark Matter Particle Physics and CSSM,
Department of Physics, Adelaide University, Adelaide, SA 5005, Australia \\
anthony.williams@adelaide.edu.au,
\orcidlink{0000-0002-1472-1592} 
ORCID 0000-0002-1472-1592
}%



\date{\today}

\begin{abstract}
The conventional strong $CP$ problem arises from the tension between
an a priori unrestricted physical parameter $\bar\theta$ and the
experimental constraint $\lvert\bar\theta\rvert\lesssim10^{-10}$.
In the standard formulation, a global topological classification of
gauge-field configuration space leads to integer-valued sectors,
$\theta$-vacua, and the interpretation of $\theta$ as a physical
vacuum angle. Alternatively, a physical flavor-singlet pseudoscalar
phase may be introduced through the quark mass matrix. The axial
anomaly relates these descriptions through the invariant combination
$\bar\theta=\theta+\arg\det M$. We ask the logically prior question
of whether known physical principles or observables require the
introduction of an independent physical flavor-singlet $CP$-violating
parameter in QCD.

We carefully distinguish consequences of local gauge invariance and
causal locality from those requiring additional global assumptions or
the introduction of an independent physical flavor-singlet
$CP$-violating parameter. For constant $\theta$, the topological
charge density is a total four-divergence with no local variational
response to compactly supported field variations, and a nonzero
$\theta$ coupling is not required by perturbative renormalization.
No established QCD phenomenology appears to require the introduction
of a physical $\bar\theta$ parameter. In the absence of such a
parameter, the topological susceptibility, axial anomaly, 't~Hooft
vertex, Leutwyler--Smilga and Witten--Veneziano relations, standard
lattice QCD results, semiclassical instanton solutions, and
established phenomenological successes of QCD all remain.

None of this prevents the introduction of $\bar\theta$ as an
independent physical parameter. The resulting conventional
formulation, with its $\theta$-vacua, topological sectors, and strong
$CP$ problem, is also entirely consistent with these same established
QCD results. In this sense, establishing that the strong $CP$ problem
is an \emph{unavoidable} feature of QCD requires showing why a
physical $\bar\theta$ parameter is \emph{required}, rather than
merely allowed.
\end{abstract}


\maketitle

\section{Introduction}
\label{Sec:Introduction}

Quantum Chromodynamics (QCD) is the non-Abelian gauge theory of the
strong interaction and forms a central part of the Standard Model of
particle physics. The most general local gauge-invariant Lagrangian
density consistent with Lorentz invariance allows a $CP$-violating term,
\begin{align}
\mathcal{L}_\theta
=
\theta (g^2/32\pi^2)F_{\mu\nu}^a\tilde{F}^{a\mu\nu}
=
\theta (g^2/16\pi^2)\tr(F_{\mu\nu}\tilde{F}^{\mu\nu}),
\end{align}
where $F_{\mu\nu}^a$ is the gluon field strength, $\tilde{F}^{a\mu\nu}$
its dual, $g$ is the strong coupling constant, and $\theta$ is a real
parameter. In the conventional formulation, the physically invariant
$CP$-violating parameter is
\begin{align}
\bar\theta=\theta+\arg\det M ,
\end{align}
where $M$ is the quark mass matrix. This parameter is, \emph{a priori},
unrestricted, whereas experimental limits on the neutron electric
dipole moment imply the stringent bound
$\lvert\bar\theta\rvert\lesssim10^{-10}$
~\cite{Baker2006ts,Abel2020gbr}. This apparent fine tuning constitutes
the conventional strong $CP$ problem
~\cite{Witten:1979vv,Crewther1979,Abel2020gbr,Vafa:1984xg}.

There are two logically distinct lines of inquiry concerning this
problem. One line of inquiry assumes a globally classified topological
structure of gauge-field configuration space and investigates the
physical consequences that follow. Within this framework one is
naturally led to topological sectors, $\theta$ vacua, and the
conventional strong $CP$ problem, together with proposed resolutions
involving, for example, axions, discrete symmetries, parity, or more
general ultraviolet constructions. This program has generated a large
and important literature, including for example
Refs.~\cite{Peccei:1977hh,Nelson:1983zb,Barr:1984qx,
Benabou2025,Kaplan:2025JHEP050,Dvali2025,Strumia2025,
Nakamura:2021meh,Schierholz:2024var,Gattringer:2020mbf,
Iida:2024irv,Kotov_2025,bonanno2025strongcpproblemtheta,
Sannino:2026wgx,Ai:2020ptm,Ai:2024vfa,Strocchi:2024tis,
Albandea:2024fui,Khoze:2025auv,Bhattacharya:2025qsk}.
For a recent discussion emphasizing the formulation of strong $CP$
violation directly in terms of the partition function and free energy,
see Ref.~\cite{Ringwald:2026apz}.

A logically distinct and prior question is whether established physical
principles or observables require an independent physical
flavor-singlet $CP$-violating parameter to be introduced in QCD.
Such a parameter may arise through the conventional global topological
construction, in which $\theta$ is interpreted as a physical vacuum
angle, or through the introduction of a physical flavor-singlet
pseudoscalar phase in the quark mass matrix; the axial anomaly relates
these descriptions through the invariant combination
$\bar\theta=\theta+\arg\det M$. The present work addresses whether
either route is required by established QCD physics. The issue
considered, for example, is not whether instantons provide an adequate
semiclassical description of QCD. They certainly do,
independently of whether or not $\bar\theta$ is introduced as
an independent physical parameter. Rather, the question being
asked is what might \emph{require} such
a parameter to be present in a consistent formulation of QCD in the
first place.

Every physical measurement is performed within a finite region of spacetime. 
Such measurements can be represented by probes of compact support, without
requiring information about the behavior of the fields at spacetime
infinity. Physical observables relevant to the present discussion are
correspondingly determined by probes of local gauge-invariant operators
within such finite regions. In particular, the local topological charge
density,
\begin{align}
q(x)
=
(g^2/32\pi^2)
F_{\mu\nu}^a\tilde{F}^{a\mu\nu},
\end{align}
is a local gauge-invariant operator, and quantities derived from its
correlation functions, such as the topological susceptibility, are
physical observables. Their existence does not, by itself, establish
that gauge-field configurations must be globally classified into
integer-valued topological sectors.

The conventional topological classification is obtained by
introducing additional assumptions concerning the global behavior of
gauge fields. These include restricting attention to sufficiently
smooth finite-action gauge configurations, requiring that the gauge
field approaches a pure gauge at spacetime infinity, and assuming
sufficiently smooth maps
\begin{align}
S^3\rightarrow SU(3),
\end{align}
which are classified by the homotopy group
$\pi_3(SU(3))=\mathbb{Z}$.
On the lattice, after taking the continuum limit and restricting
attention to sufficiently smooth gauge configurations, the
analogous topological construction is based on
$T^4\rightarrow SU(3)$. These assumptions are sufficient to obtain the
familiar decomposition of gauge-field configuration space into
integer-valued topological sectors and thereby construct the
conventional $\theta$-vacuum framework. This framework is mathematically
consistent and has proved remarkably successful in the study of
nonperturbative QCD. The question addressed here is not whether this
construction is permissible or internally consistent, but whether
established QCD physics requires it.

\begin{figure}[htbp]
    \centering
    \includegraphics[width=0.68\textwidth]{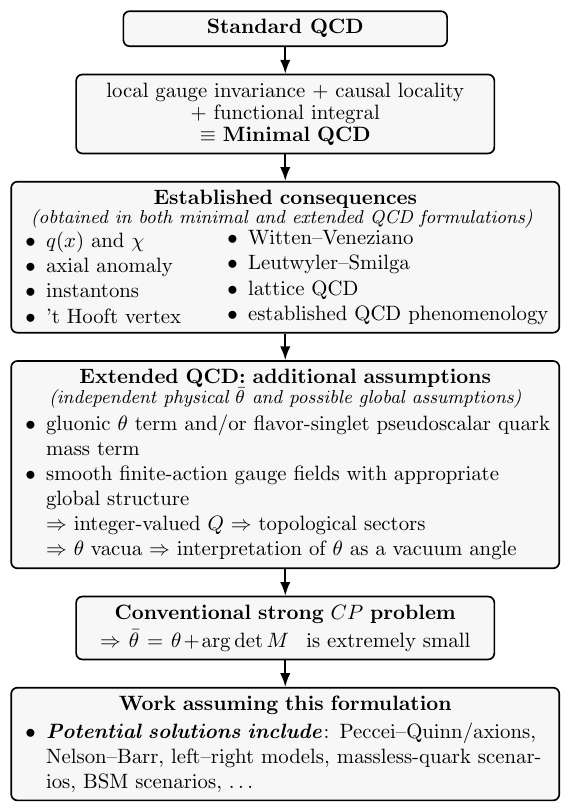}
    \caption{
Illustration of the logical structure discussed in this work.
Standard $CP$-conserving QCD together with local gauge invariance,
causal locality, and the functional integral defines what we refer to
here as \emph{minimal QCD}, in which no independent physical
flavor-singlet $CP$-violating parameter $\bar\theta$ is introduced.
A broad class of established nonperturbative results follows within
this framework. We refer to formulations in which an independent
physical $\bar\theta$ parameter is introduced as \emph{extended QCD}.
The same physical $\bar\theta$ may be represented by a gluonic
$\theta$ term, a flavor-singlet pseudoscalar phase of the quark mass
matrix, or a combination of the two. Additional global assumptions
concerning smoothness, boundary conditions, and topology lead to the
conventional $\theta$-vacuum construction, with topological sectors
and the interpretation of $\theta$ as a periodic vacuum angle. It is
within extended QCD that the conventional strong $CP$ problem arises.
}
    \label{Fig:two_forms_fig.pdf}
\end{figure}

For the purposes of this paper, for convenience,
we shall refer to the minimal set of assumptions
consisting of the standard $CP$-conserving QCD Lagrangian, local gauge
invariance, causal locality, and the functional integral over gauge
fields as \emph{minimal QCD}. In particular, no independent physical
flavor-singlet $CP$-violating parameter $\bar\theta$ is introduced.
We shall refer to formulations in which such an independent physical
parameter is introduced as \emph{extended QCD}. The physical parameter
may be represented by a gluonic $\theta$ term, a flavor-singlet
pseudoscalar phase in the quark mass matrix, or a combination of the
two. Since an anomalous axial transformation can transfer the phase
between the gluonic and mass terms, only the invariant combination
$\bar\theta=\theta+\arg\det M$ is physical. Additional global
assumptions concerning smoothness, boundary conditions, and topology
lead, in particular, to the conventional $\theta$-vacuum construction
and the interpretation of $\theta$ as a periodic vacuum angle. For
brevity, we shall refer collectively to assumptions or ingredients
that introduce an independent physical $\bar\theta$ parameter as the
\emph{additional assumptions}.

It should be emphasized that the present work does not argue that an
independent $CP$-violating term proportional to $q(x)$ is incompatible
with locality, Lorentz invariance, gauge invariance, or renormalizability.
Such a term is certainly allowed by these symmetries. However, it is
important to note that for the conventional
constant $\theta$ the topological density is a total four-divergence,
$q(x)=\partial_\mu K^\mu(x)$, so that for field variations of compact
support
\begin{align}
\textstyle \delta\left[\theta\int d^4x\,q(x)\right]
=\theta\int d^4x\,\partial_\mu\delta K^\mu(x) =0 .
\end{align}
Consequently, a constant $\theta$ term produces no local variational
response, is invisible in ordinary perturbation theory, and perturbative
renormalization does not require a nonzero $\theta$ coupling to be
introduced. It is therefore distinct in important respects from
ordinary symmetry-allowed interactions whose inclusion may be required
for closure of the renormalized theory. The fact that the operator
$q(x)$ is allowed by the local symmetries consequently establishes
that an independent $\theta$ coupling may be introduced, but
does not by itself establish that such a coupling must be
introduced. The question addressed here is what, if anything,
requires an independent physical flavor-singlet $CP$-violating
parameter, whether represented by $\theta$ and/or by a pseudoscalar phase
of the quark mass matrix, rather than merely \emph{allowing} one. This
distinction is central to the strong $CP$ problem, since the
conventional fine-tuning problem presupposes that an independent
physical $CP$-violating parameter is an \emph{unavoidable} feature of a
consistent formulation of QCD.

We therefore begin without assuming either a global decomposition of
gauge-field configuration space into integer-valued topological sectors
or an independent physical $CP$-violating parameter. Physical
observables are defined through correlation functions of local
gauge-invariant operators. We first establish which familiar
nonperturbative results follow within this framework and then identify
the additional assumptions required to obtain the conventional global
topological formulation.

Our purpose is not to question the mathematical consistency or
phenomenological success of extended QCD or of its conventional
global topological formulation. An independent physical
$\bar\theta$ parameter may consistently be introduced, and the
conventional formulation, including its $\theta$ vacua, topological
sectors, semiclassical instanton solutions and their conventional
interpretation, and strong $CP$ problem, is entirely legitimate.
We ask instead whether established QCD physics requires this choice.
As we show below, the established local and nonperturbative results
considered here remain in the absence of an independent physical
$\bar\theta$ parameter. Establishing the conventional strong $CP$
problem as an \emph{unavoidable} feature of QCD therefore requires
showing why such a physical parameter is \emph{required}, rather than
merely \emph{allowed}.

The present analysis is accordingly compatible with studies of neutron
electric dipole moments, axion and Peccei--Quinn models, Nelson--Barr
constructions, lattice topology, and other investigations performed
within the conventional framework. Such studies investigate the
consequences of a physical $\bar\theta$ or mechanisms that account for
its observed smallness. Our question is logically prior: what
established physical principle or observable requires the independent
physical $CP$-violating parameter upon which the conventional strong
$CP$ problem is predicated?

The paper is organized as follows. In Sec.~II we introduce $\theta$ as
a source parameter in the generating functional and derive the
topological susceptibility and related quantities without assuming a
global topological classification of gauge fields or that $\theta$ is
an independent physical parameter. We then discuss several standard
nonperturbative results that likewise do not require such a
classification. In Sec.~III we introduce the additional assumptions
leading to integer-valued topological sectors and the conventional
$\theta$-vacuum construction. Finally, Sec.~IV summarizes our
conclusions.

\section{Minimal QCD: Standard Nonperturbative Results}

In this section we establish standard nonperturbative
results within the framework of minimal QCD. These
results follow from local gauge invariance, causal locality,
and the functional integral, without introducing an independent
physical flavor-singlet $CP$-violating parameter or the additional
global assumptions leading to a topological classification of
gauge-field configurations.

\subsection{QCD and Generating Functionals}
\label{Sec:local_QCD}

Working in natural units, $\hbar=c=1$, the QCD action in Minkowski space is
\begin{align}
S_M=\textstyle \int\! d^4x\,\mathcal{L}_M,
\end{align}
where
\begin{align}\label{Eq:L_M}
\mathcal{L}_M
=
-\textstyle \frac12\tr(F_{\mu\nu}F^{\mu\nu})
+\bar\psi(i\slashed{D}-m)\psi .
\end{align}
Here $m=\operatorname{diag}(m_u,m_d,\ldots)\in\mathbb{R}$ denotes the 
real diagonal quark mass matrix, which may always
be chosen diagonal by the allowed flavor field redefinitions. Thus no
independent flavor-singlet pseudoscalar mass term, and hence no
independent physical $CP$-violating parameter $\bar\theta$, is
introduced in minimal QCD.
This is part of the definition of minimal QCD adopted in
Sec.~\ref{Sec:Introduction}.
Throughout this work we use the notation and conventions of
Ref.~\cite{Williams:2022bzq}.
After Wick rotation, $x^0\rightarrow-ix_4$, the Euclidean action becomes
\begin{align}
S_E=\textstyle \int \! d^4x_E\,\mathcal{L}_E,
\end{align}
with
\begin{align}\label{Eq:L_E}
\mathcal{L}_E
=
\textstyle \frac12\tr(F_{\mu\nu}F_{\mu\nu})
+\bar\psi(\slashed{D}_E+m)\psi .
\end{align}
The corresponding generating functionals are
\begin{align}\label{Eq:QCD_genZ}
Z_M
&= \textstyle \int\!
\mathcal{D}A_\mu\,
\mathcal{D}\bar\psi\,
\mathcal{D}\psi\;
e^{\,iS_M[A,\psi,\bar\psi]}
\, , \qquad
Z_E = \textstyle \int \!\mathcal{D}A_\mu\,
\mathcal{D}\bar\psi\,
\mathcal{D}\psi\;
e^{-S_E[A,\psi,\bar\psi]} .
\end{align}
The Euclidean functional integral is formally obtained by Wick
rotation from the Minkowski theory and provides the standard
formulation used in nonperturbative analyses, including lattice QCD.

At this stage no assumption has been made concerning the global
smoothness or topological classification of gauge-field
configurations. The functional integral is written formally over gauge
fields, and the additional global assumptions leading to the conventional
topological formulation will be introduced only in
Sec.~\ref{Sec:Global_Classification}.

For any gauge-invariant observable
$\hat O\equiv O[\hat A,\hat{\bar\psi},\hat\psi]$,
its vacuum expectation value is
\begin{align}\label{Eq:QCD_observ}
\langle\hat O\rangle_M
&= \frac{ \displaystyle \textstyle \int\! \mathcal{D}A_\mu\,
\mathcal{D}\bar\psi\, \mathcal{D}\psi\; O[A,\bar\psi,\psi]\,
e^{\,iS_M}
}{
\displaystyle
\textstyle\int \! \mathcal{D}A_\mu\, \mathcal{D}\bar\psi\,
\mathcal{D}\psi\; e^{\,iS_M}
}
\, , \qquad
\langle\hat O\rangle_E =
\frac{ \displaystyle \textstyle \int\! \mathcal{D}A_\mu\,
\mathcal{D}\bar\psi\, \mathcal{D}\psi\; O[A,\bar\psi,\psi]\,
e^{-S_E}
}{
\displaystyle \textstyle \int \!\mathcal{D}A_\mu\,
\mathcal{D}\bar\psi\, \mathcal{D}\psi\; e^{-S_E} }.
\end{align}

The topological charge is defined by
\begin{align}\label{Eq:top_Q}
Q
&=
{\textstyle \int} d^4x_M\,q_M(x)
=
\frac{g^2}{16\pi^2}
{\textstyle \int} d^4x_M\,
\tr(F_{\mu\nu}\tilde F^{\mu\nu})
\nonumber\\
&=
{\textstyle \int} d^4x_E\,q_E(x)
=
\frac{g^2}{16\pi^2}
{\textstyle \int} d^4x_E\,
\tr(F_{\mu\nu}\tilde F_{\mu\nu}),
\end{align}
where $q_M(x)$ and $q_E(x)$ denote the Minkowski and Euclidean
topological charge densities, respectively. The quantity $Q$ is the same
in both formulations~\cite{Williams:2026cec}.

The topological charge density may be written locally as
\begin{align}\label{Eq:dmuCS=q}
q(x)
=
\frac{g^2}{32\pi^2}
F_{\mu\nu}^a\tilde F^{a\mu\nu}
=
\partial_\mu K^\mu(x),
\end{align}
where $K^\mu(x)$ is the Chern--Simons current,
\begin{align}
K^\mu
=
\frac{g^2}{16\pi^2}
\epsilon^{\mu\nu\rho\sigma}
\left(
A_\nu^a\partial_\rho A_\sigma^a
+
\frac13
g f^{abc}
A_\nu^aA_\rho^bA_\sigma^c
\right).
\end{align}
Although $K^\mu(x)$ is neither gauge invariant nor globally defined,
the identity $q(x)=\partial_\mu K^\mu(x)$ holds locally within any
coordinate patch on which a gauge potential is defined.

Under an infinitesimal variation of the gauge field,
\begin{align}
\delta S_\theta
&=
\theta
{\textstyle \int} d^4x\,\delta q(x)
=
\theta
{\textstyle \int} d^4x\,
\partial_\mu[\delta K^\mu(x)] .
\end{align}
For compactly supported variations of the gauge field,
$\delta K^\mu$ is also compactly supported and therefore
\begin{align}\label{Eq:int_dmuK}
\delta S_\theta
=
\theta
{\textstyle \int} d^4x\,
\partial_\mu[\delta K^\mu(x)]
=
{\textstyle \int}_{\partial\mathcal M}
d\Sigma_\mu\,
\delta K^\mu
=
0.
\end{align}
Consequently, the $\theta$ term contributes neither to the
Euler--Lagrange equations of motion nor, when perturbation theory is
constructed about the trivial vacuum using compactly supported
fluctuations, to the quadratic action determining propagators or to the
interaction vertices. It therefore does not appear in perturbative
Feynman rules.

Define the Minkowski and Euclidean Lagrangian densities including a
$\theta$ term by
\begin{align}\label{Eq:L_thetas}
\mathcal{L}_M(\theta)
&\equiv
\mathcal{L}_M+\theta q_M
=
-\frac12\tr(F_{\mu\nu}F^{\mu\nu})
+\bar\psi(i\slashed{D}-m)\psi
+\theta\frac{g^2}{16\pi^2}
\tr(F_{\mu\nu}\tilde F^{\mu\nu}),
\\
\mathcal{L}_E(\theta)
&\equiv
\mathcal{L}_E-i\theta q_E
=
\frac12\tr(F_{\mu\nu}F_{\mu\nu})
+\bar\psi(\slashed{D}_E+m)\psi
-i\theta\frac{g^2}{16\pi^2}
\tr(F_{\mu\nu}\tilde F_{\mu\nu}) .
\nonumber
\end{align}

The corresponding generating functionals are
\begin{align}\label{Eq:QCD_genZ_theta}
Z_M(\theta)
&=
{\textstyle \int}
\mathcal{D}A_\mu\,
\mathcal{D}\bar\psi\,
\mathcal{D}\psi\;
e^{\,iS_M(\theta)}
\, , \qquad
Z_E(\theta)
=
{\textstyle \int}
\mathcal{D}A_\mu\,
\mathcal{D}\bar\psi\,
\mathcal{D}\psi\;
e^{-S_E(\theta)} .
\end{align}
Throughout this section, the parameter $\theta$ is introduced
solely as the \emph{source parameter} of the generating functional.
Its introduction as a source does not enlarge minimal QCD by an
independent physical $CP$-violating parameter, just as the introduction
of a source for any local composite operator does not make that source
a physical coupling of the theory. No interpretation of $\theta$ as
a physical vacuum angle is assumed.
Differentiation with respect to $\theta$ therefore 
generates moments and correlation
functions of the integrated topological charge,
\begin{align}
Q
=
{\textstyle \int} d^4x\,q(x).
\end{align}

One could equally well introduce a spacetime-dependent source
$\theta(x)$ in the generating functional $Z_E[\theta]$ and obtain
correlation functions of the local topological charge density
$q(x)$ by functional differentiation. For the purposes of the
present work, however, this additional generality is unnecessary,
since we require only the spacetime-integrated correlation functions obtained
from differentiation with respect to a constant source. We therefore
treat $\theta$ as a real constant throughout.

Since the functional integral at $\theta=0$ sums over every gauge-field
configuration together with its $CP$ conjugate, and since
$Q\rightarrow-Q$ under $CP$, the contributions proportional to odd
powers of the source $\theta$ cancel pairwise. Consequently,
\begin{align}
Z_M(\theta)=Z_M(-\theta),
\qquad
Z_E(\theta)=Z_E(-\theta),
\end{align}
and all odd derivatives of $Z(\theta)$ and $\ln Z(\theta)$ vanish at
$\theta=0$. In the Euclidean formulation, the paired contributions
also make $Z_E(\theta)$ real.
The topological susceptibility therefore need not vanish
at $\theta=0$; rather, it is determined by the second derivative of
the generating functional with respect to the source parameter
$\theta$. This observation forms the starting point for the discussion
in the following subsection.

If, in addition, $Q$ is assumed to be integer-valued, then the same
generating functional is periodic under $\theta\rightarrow\theta+2\pi$,
since
\begin{align}
e^{i(\theta+2\pi)Q}=e^{i\theta Q}
\qquad
(Q\in\mathbb Z).
\end{align}
This periodicity is therefore not a consequence of the local operator
$q(x)$ alone, but follows only after the additional global assumptions
leading to integer-valued topological charge have been introduced.
This point will be discussed in Sec.~\ref{Sec:Global_Classification}.

\subsection{Topological susceptibility}
\label{Sec:top_suscept}

The topological susceptibility plays a central role in
nonperturbative QCD. It is defined as the zero-momentum two-point
correlation function of the local topological charge density and
measures fluctuations of the integrated topological charge in the
vacuum. In Minkowski space,
\begin{align}
\chi
=
\textstyle
\int d^4x_M\,
\langle\Omega|
T\,\hat q_M(x)\hat q_M(0)
|\Omega\rangle ,
\end{align}
where
$\ket{\Omega}\equiv\ket{\Omega}_{\theta=0}$.
Since Euclidean correlation functions do not require time ordering,
the corresponding definition is
\begin{align}\label{Eq:chi_defn}
\chi
=
\textstyle
\int d^4x_E\,
\langle\Omega|
\hat q_E(x)\hat q_E(0)
|\Omega\rangle .
\end{align}

Consider a finite Euclidean four-volume $V$ with periodic boundary
conditions. This corresponds to the continuum limit of the four-torus
employed in lattice QCD. Defining the operator
\begin{align}
\hat Q
=
\textstyle
\int d^4x_E\,\hat q_E(x),
\end{align}
it follows that
\begin{align}
\hat Q^2
=
\textstyle
\int d^4x_E
\int d^4y_E\,
\hat q_E(x)\hat q_E(y).
\end{align}
At this stage $\hat Q$ denotes only the spacetime integral of the
local operator $\hat q_E(x)$. No assumption is being made that this
operator has integer-valued eigenvalues or that it labels
topological sectors of gauge-field configuration space.
Taking the Euclidean vacuum expectation value at $\theta=0$ gives
\begin{align}
\langle\hat Q^2\rangle
&=
\textstyle
\int d^4x_E
\int d^4y_E\,
\langle
\Omega|
\hat q_E(x)\hat q_E(y)
|\Omega
\rangle
\nonumber\\
&=
V\,
\textstyle
\int d^4x_E\,
\langle
\Omega|
\hat q_E(x)\hat q_E(0)
|\Omega
\rangle
=
V\chi ,
\nonumber
\end{align}
where translational invariance has been used in the second line.
Consequently,
\begin{align}\label{Eq:chi_Q2}
\chi
=
\frac{\langle\hat Q^2\rangle}{V}
=
\textstyle
\int d^4x_E\,
\langle
\Omega|
\hat q_E(x)\hat q_E(0)
|\Omega
\rangle .
\end{align}
After the usual regularization of the contact singularity at
$x=0$, the topological susceptibility is a
renormalization-group-invariant observable. This definition depends
only on the local operator $\hat q(x)$ and its correlation function.

For QCD at $\theta=0$ with the physical quark masses chosen positive,
the vacuum $\ket{\Omega}$ is $CP$ invariant. Since the topological
charge density is $CP$ odd,
\begin{align}
\langle\hat q(x)\rangle=0,
\qquad
\langle\hat Q\rangle=0,
\end{align}
and therefore
\begin{align}
\chi
=
\frac{\langle\hat Q^2\rangle}{V}
=
\frac{\langle\hat Q^2\rangle-\langle\hat Q\rangle^2}{V},
\end{align}
so that the topological susceptibility is simply the variance density
of the integrated topological charge.

For general $\theta$ we correspondingly define
\begin{align}
\chi_\theta
=
\frac{
\langle\hat Q^2\rangle_\theta
-
\langle\hat Q\rangle_\theta^2
}{V},
\end{align}
where
\begin{align}
\langle\cdots\rangle_\theta
\equiv
{}_E\!\langle\Omega_\theta|
\cdots
|\Omega_\theta\rangle_E .
\end{align}
Here the subscript $\theta$ denotes expectation values evaluated in
the generating functional with source parameter $\theta$. No
interpretation of $\theta$ as a physical vacuum angle, or of the
corresponding functional integral in terms of a superposition of
integer-valued topological sectors, is assumed at this stage.
We shall show in Sec.~III that, once the additional global assumptions
underlying the conventional topological formulation are adopted, this
ground state is naturally identified with the conventional $\theta$
vacuum.

Consider now the Euclidean generating functional including the
$\theta$ term, $Z_E(\theta)$, defined in
\Eq{Eq:QCD_genZ_theta}. Then
\begin{align}\label{Eq:chi_theta_V}
Z_E(\theta)
&=
\textstyle
\int
\mathcal{D}A_\mu
\mathcal{D}\bar\psi
\mathcal{D}\psi\,
e^{-\left[S_E-i\theta\int d^4z_E\,q_E(z)\right]},
\nonumber\\
\frac{\partial\ln Z_E}{\partial\theta}
&=
\frac{1}{Z_E(\theta)}
\frac{\partial Z_E(\theta)}{\partial\theta}
=
i\langle\hat Q\rangle_\theta ,
\nonumber\\
\frac{\partial^2\ln Z_E}{\partial\theta^2}
&=
\frac{1}{Z_E(\theta)}
\frac{\partial^2Z_E(\theta)}{\partial\theta^2}
-
\left(
\frac{1}{Z_E(\theta)}
\frac{\partial Z_E(\theta)}{\partial\theta}
\right)^2
=
-
\langle\hat Q^2\rangle_\theta
+
\langle\hat Q\rangle_\theta^2
=
-\chi_\theta V .
\end{align}
Every step in these identities follows directly from the definition of
the generating functional and the local operator $q(x)$. Up to this
point no assumption has been made that the integrated topological
charge $Q$ is integer valued or that gauge-field configurations are
globally classified into topological sectors.

Since $Z_E(\theta)$ is an even function of $\theta$,
$\langle\hat Q\rangle_\theta$ is an odd function of $\theta$, and so
\begin{align}
\langle\hat Q\rangle_{\theta=0} = \braket{\hat Q} =0.
\end{align}
Similarly, $\chi_\theta$ is an even function of $\theta$. Hence
$\theta=0$ implies $\chi_\theta=\chi$, but it does \emph{not} imply
that $\chi=\langle\hat Q^2\rangle/V=0$. Although the topological
susceptibility is defined as a function of the source parameter
$\theta$, its value at $\theta=0$ is generally nonzero.
Consequently, the existence of a nonvanishing topological
susceptibility does not by itself imply the existence of an independent
physical vacuum angle.

On a Euclidean four-torus of spatial volume $V_3$ and Euclidean time
extent $T_E$, the four-volume is $V=V_3T_E$ and the Euclidean
generating functional may be identified with the partition function,
\begin{align}
Z_E(\theta)
=
\mathrm{Tr}( e^{-\hat H_\theta T_E})
=
{\textstyle \sum_n}
e^{-E_n(\theta)T_E},
\end{align}
where $\hat H_\theta$ is the Hamiltonian corresponding to the
$\theta$-dependent action. In the limit $T_E\rightarrow\infty$, the
ground state dominates,
\begin{align}\label{Eq:vareps_0}
-\frac{\ln Z_E(\theta)}{V}
\xrightarrow[T_E\rightarrow\infty]{}
\frac{E_0(\theta)}{V_3}
\equiv
\varepsilon_0(\theta),
\end{align}
where $\varepsilon_0(\theta)$ denotes the corresponding ground-state
energy density associated with the source parameter $\theta$.

Combining \Eq{Eq:chi_theta_V} and \Eq{Eq:vareps_0} therefore gives
\begin{align}\label{Eq:chi_vareps}
\chi_\theta
=
\frac{\partial^2\varepsilon_0(\theta)}
{\partial\theta^2}
=
\frac{
\langle\hat Q^2\rangle_\theta
-
\langle\hat Q\rangle_\theta^2
}{V}.
\end{align}
Thus the topological susceptibility is the second derivative of the
vacuum energy density with respect to the source parameter $\theta$,
or equivalently the variance density of the integrated topological
charge. This derivation relies only on the local topological charge
density, the generating functional, and the source parameter
$\theta$. It does not require the additional global assumptions leading
to an integer-valued topological classification of gauge fields.

\subsection{Semiclassical approximation and instantons}
\label{Sec.semiclass}

It is important to distinguish the exact quantum functional integral
from the classical and semiclassical descriptions derived from it.
Macroscopic observables do not probe the microscopic field
configuration pointwise, but only fields smeared over finite spacetime
regions whose dimensions are large compared with microscopic
correlation lengths. For example, defining the coarse-grained field
\begin{align}
\phi_R
=
\frac{1}{V_R}\int_R d^4x\,\phi(x),
\end{align}
a macroscopic observable such as
\begin{align}
O_R[\phi]
=
\phi_R^2
=
\left[
\frac{1}{V_R}\int_R d^4x\,\phi(x)
\right]^2
\end{align}
depends only on the coarse-grained behavior of the field over the
region $R$, where $V_R$ is its spacetime volume. The effective
macroscopic dynamics is correspondingly formulated in terms of such
coarse-grained variables rather than the microscopic field
configurations themselves.

In Euclidean field theory, when the correlation length is finite,
fluctuations in sufficiently widely separated subregions of $R$ are
approximately uncorrelated. A coarse-grained field therefore receives
contributions from many nearly independent microscopic degrees of
freedom. Consequently, deviations of a macroscopic coarse-grained
configuration from its typical value are increasingly suppressed as
the coarse-graining volume grows, schematically as
\begin{align}
P(\phi_R)
\sim
\exp[-V_R I(\phi_R)/\hbar],
\qquad
I\geq0,
\end{align}
illustrating the concentration of measure
\cite{ZinnJustin2002,Brezis2011,Varadhan1984}. In the macroscopic
limit, the probability distribution of the coarse-grained fields is
therefore sharply concentrated about the stationary configurations of
the appropriate coarse-grained effective action. Classical or
semiclassical dynamics can consequently provide an accurate
description of macroscopic fields even though the exact microscopic
functional integral is supported on rough, distributional field
configurations.

Instantons are smooth finite-action stationary points of the Euclidean
Yang--Mills action that arise naturally within this semiclassical
description. For the usual smooth finite-action configurations, the
field strength vanishes asymptotically and the gauge field approaches
a pure gauge at spacetime infinity. These conditions define the
classical saddle-point problem and are sufficient to identify
instanton solutions and organize a semiclassical expansion about
them.

The assumptions used to define the semiclassical saddle-point problem
should not, however, be confused with assumptions concerning the exact
functional integral. Semiclassical approximations identify stationary
configurations about which a given functional integral may be
approximated, but they do not determine the space of gauge-field
configurations over which the exact functional integral is defined.
Thus, although the boundary and smoothness conditions used to
construct instantons are sufficient to obtain the conventional
topological classification of these classical solutions, they do not
by themselves establish that the exact quantum functional integral
must be formulated over gauge fields admitting a global
integer-valued topological classification. Instantons therefore
provide a powerful semiclassical organizing principle without, by
themselves, establishing that the associated global topological
framework is required to account for the established physical
observables associated with QCD.

The same distinction applies to the 't~Hooft effective interaction.
Its semiclassical derivation uses the fermion zero modes of a smooth
instanton background and provides a particular semiclassical
realization of the anomalous $U(1)_A$ dynamics. The resulting
interaction is expressed in terms of local fermionic operators. Its
semiclassical derivation therefore relies on the existence of the
instanton saddle point, but does not by itself establish that the
exact gauge-field measure must be regarded as fundamentally
decomposed into globally classified integer-valued topological
sectors. The same distinction also underlies the
Witten--Veneziano relation, which is discussed separately in
Sec.~\ref{Sec:WV_formula}.

\subsection{Axial anomaly and the emergence of $\bar\theta$}
\label{Sec:theta_bar}

We first consider the most general quark mass matrix before examining
how the axial anomaly leads to the invariant parameter
$\bar\theta$. The quark mass term in the QCD Lagrangian may be written
as
\begin{align}\label{Eq:quark_L_m}
\mathcal{L}_m
=
-\bar\psi_R M\psi_L
-\bar\psi_L M^\dagger\psi_R ,
\end{align}
where $\psi$ is an $N_f$-component Dirac spinor in flavor space and
$M$ is a general complex $N_f\times N_f$ matrix. Equivalently, the
most general Lorentz-invariant Hermitian mass term may be written as
\begin{align}\label{Eq:theta_as_mass}
\mathcal{L}_m
=
-\bar\psi
\left(
M_S+i\gamma_5M_P
\right)
\psi,
\end{align}
where $M_S$ and $M_P$ are Hermitian matrices in flavor space. Thus a
general complex quark mass matrix contains both scalar and
pseudoscalar mass couplings, the latter allowing explicit $CP$
violation in the fermion mass sector.

Any complex matrix admits a biunitary (singular-value)
decomposition,
\begin{align}
M
=
U_RDU_L^\dagger ,
\qquad
D=\mathrm{diag}(m_1,\ldots,m_{N_f}),
\end{align}
where $U_L,U_R\in U(N_f)$ and the singular values $m_i$ are real and
nonnegative. From a purely algebraic viewpoint, therefore, every
complex mass matrix may be brought to a real diagonal form.

In quantum field theory, however, it is essential to distinguish this
algebraic statement from the effect of the corresponding field
redefinitions on the quantum functional measure. Locally near the
identity, the chiral flavor group may be written as
\begin{align}
U(N_f)_L\times U(N_f)_R
&\simeq
SU(N_f)_L\times SU(N_f)_R
\times U(1)_L\times U(1)_R
\nonumber\\
&\simeq
SU(N_f)_L\times SU(N_f)_R
\times U(1)_V\times U(1)_A,
\end{align}
where the global quotient by the common discrete center is irrelevant
for the present local argument. In the quantum theory the $U(1)_A$
subgroup is anomalous: an axial change of fermion variables produces
a nontrivial Jacobian in the functional measure. Consequently, the
non-anomalous transformations
\begin{align}
SU(N_f)_L
\times
SU(N_f)_R
\times
U(1)_V
\end{align}
may be used to simplify the mass matrix without inducing an additional
term in the action, whereas a $U(1)_A$ transformation simultaneously
changes the coefficient of the local operator $F\tilde F$ through
the anomalous Jacobian.

Writing
\begin{align}
U_R=e^{i\phi_R/N_f}V_R,
\qquad
U_L=e^{i\phi_L/N_f}V_L,
\end{align}
with $V_R,V_L\in SU(N_f)$, one finds
\begin{align}
V_R^\dagger M V_L
=
e^{i\Phi/N_f}D,
\qquad
\Phi=\phi_R-\phi_L.
\end{align}
Thus the non-anomalous transformations remove all relative phases
from the quark mass matrix while leaving a single flavor-singlet
phase associated with the anomalous $U(1)_A$ transformation. The
common phase corresponding to $U(1)_V$ drops out of the mass matrix
entirely. Since $\det D$ is real and positive when $\det M\neq0$,
\begin{align}
\Phi
=
\arg\det M
\qquad
(\mathrm{mod}\ 2\pi).
\end{align}
Thus, after exhausting the non-anomalous transformations, the most
general nonsingular quark mass matrix contains a single irreducible
flavor-singlet phase proportional to $\arg\det M$. Its physical
significance is determined by the axial anomaly.

Consider the global axial change of fermion variables
\begin{align}
\psi
&\rightarrow
e^{i\alpha\gamma_5}\psi,
\qquad
\bar\psi
\rightarrow
\bar\psi\,e^{i\alpha\gamma_5}.
\end{align}
Classically, the massless QCD action is invariant under this
transformation. In the quantum theory, however, the fermion
functional measure is not invariant. The local anomalous Jacobian is
equivalently represented, whenever its spacetime integral is
well-defined, by
\begin{align}
\mathcal{D}\bar\psi\,
\mathcal{D}\psi
\rightarrow
\mathcal{D}\bar\psi\,
\mathcal{D}\psi\,
\exp\!\left(
i\,2N_f\alpha\,Q
\right),
\end{align}
where
\begin{align}
Q
=
\frac{g^2}{32\pi^2}
\int d^4x\,
F_{\mu\nu}^a
\tilde F^{a\mu\nu}.
\end{align}
No assumption that $Q$ is integer valued is required for this
anomalous change of variables. The Jacobian is equivalent to the
shift
\begin{align}
\theta
\rightarrow
\theta+2N_f\alpha .
\end{align}
Thus the anomalous transformation shifts the source parameter
$\theta$ and simultaneously rotates the quark mass matrix.

The same axial transformation multiplies the quark mass matrix by an
overall phase,
\begin{align}
M
\rightarrow
M'
=
e^{-2i\alpha}M,
\end{align}
so that
\begin{align}
\arg\det M
\rightarrow
\arg\det M
-
2N_f\alpha .
\end{align}
Consequently, the combination
\begin{align}
\bar\theta
\equiv
\theta+\arg\det M
\end{align}
is invariant under the anomalous axial transformation,
\begin{align}
\bar\theta
\rightarrow
\bar\theta.
\end{align}
Thus, for $\det M\neq0$, the flavor-singlet strong-sector
$CP$ violation associated with the source parameter $\theta$ and the
complex quark mass determinant is encoded in the single invariant
parameter
\begin{align}
\bar\theta
=
\theta+\arg\det M.
\end{align}

The axial anomaly itself is a local statement. Whenever the
corresponding integrated change of fermion variables is well-defined,
it implies the compensating transformations of the coefficient of
$F\tilde F$ and the flavor-singlet phase of the quark mass matrix
derived above. No assumption that the spacetime integral of
$F\tilde F$ is integer valued, or that the gauge-field configuration
space decomposes into sectors labeled by such an integer, is required
for this relation. One may therefore choose a field basis in which
this strong-sector $CP$ violation is represented either by a
flavor-singlet pseudoscalar quark mass term or by the source parameter
$\theta$, while the invariant combination $\bar\theta$ remains
unchanged. Any further connection between the source parameter
$\theta$ and a global topological classification of gauge-field
configurations enters only after additional assumptions concerning
the global structure of gauge-field configuration space are
introduced in Sec.~III.

It is important to distinguish two logically separate questions. The
local axial anomaly identifies the invariant combination
$\bar\theta=\theta+\arg\det M$ once the corresponding $CP$-violating
terms are admitted. This establishes the equivalence of their
parameterizations, but it does not by itself require that either an
independent $\theta$ parameter or an independent flavor-singlet
$CP$-violating quark mass term be present. Within the framework of
minimal QCD introduced in this work, neither is assumed a priori. The
question addressed here is therefore not how the two
parameterizations are related, but whether any established physical
observable requires introducing either of these additional
$CP$-violating parameters in the first place.

The original estimate of the neutron electric dipole moment by
Crewther {\it et al.}~\citep{Crewther1979} provides an important
illustration of this point. Their calculation exploits the
equivalence between the $\theta$ term and a flavor-singlet
pseudoscalar quark mass term. Although the $\theta$ term does not
modify perturbative Feynman rules, as discussed in
Sec.~\ref{Sec:local_QCD}, the induced pseudoscalar interaction
contributes through intrinsically nonperturbative chiral dynamics
involving the quark condensate, Goldstone bosons, and hadronic matrix
elements. Thus the absence of perturbative effects does not imply the
absence of observable nonperturbative consequences once such a
$CP$-violating source is introduced.

Finally, suppose one or more quark masses vanish so that
\begin{align}
\det M=0.
\end{align}
Then $\arg\det M$ is undefined. If one quark is exactly massless, an
axial redefinition of that quark field changes the coefficient of
$F\tilde F$ through the anomalous Jacobian but introduces no
compensating phase into a mass term for that quark, since no such mass
term is present. Within the conventional formulation in which a
$\theta$ source is admitted, the source parameter $\theta$ may
therefore be removed completely. The generating functional is then
independent of $\theta$, and the vacuum energy is likewise independent
of $\theta$, implying
\begin{align}
\chi_\theta
=
\frac{\partial^2\varepsilon_0(\theta)}
{\partial\theta^2}
=
0.
\end{align}
Thus, within the conventional formulation in which a $\theta$ source
is admitted, an exactly massless quark provides the well-known
exceptional case in which that source becomes physically redundant
and the conventional strong $CP$ problem does not arise. This
conclusion follows from the anomalous change of fermion variables and
does not require an integer-valued topological classification of the
gauge-field configurations.

\subsection{Witten--Veneziano formula and the $\eta'$ mass}
\label{Sec:WV_formula}

The Witten--Veneziano relation is widely regarded as one of the
strongest pieces of evidence for the physical importance of the local
topological charge density in QCD. It is therefore important to
identify precisely which assumptions enter its derivation. As we now
show, the relation relies upon the anomalous singlet axial Ward
identity, a nonvanishing pure Yang--Mills topological susceptibility,
and the large-$N_c$ expansion. It does not, by itself, require that
the exact QCD functional integral be fundamentally decomposed into
configuration-by-configuration sectors labeled by an integer-valued
topological charge.

The resolution of the $U(1)_A$ problem is conceptually independent of
the value of the source parameter $\theta$. In massless QCD the
classical Lagrangian possesses a singlet axial $U(1)_A$ symmetry, but
this is not a symmetry of the quantum theory because of the axial
anomaly. In the instanton semiclassical approximation, the anomalous
$U(1)_A$ dynamics is realized through the 't~Hooft effective
interaction generated by fermion zero modes in smooth instanton
backgrounds, as discussed in the previous subsection. More generally,
the anomalous singlet axial Ward identity is an exact local statement
of the quantum theory and does not depend upon the validity of the
semiclassical instanton approximation.

The quantity entering the Witten--Veneziano relation is the
topological susceptibility of pure Yang--Mills theory,
\begin{align}
\chi_{\rm YM}
\equiv
\int \! d^4x_E\,
\braket{
\hat q(x)\hat q(0)
}_{\rm YM},
\qquad
\hat q(x)
=
\frac{g^2}{32\pi^2}
\hat F_{\mu\nu}^a
\hat{\tilde F}^a_{\mu\nu},
\label{Eq:chiYM_def}
\end{align}
evaluated at vanishing source parameter, $\theta=0$. This should be
distinguished from the topological susceptibility of full QCD
discussed in Sec.~\ref{Sec:top_suscept}. In particular, if one or more
quarks are exactly massless, the full QCD susceptibility vanishes
because the $\theta$ dependence may be removed by an anomalous chiral
change of fermion variables, whereas $\chi_{\rm YM}$ is not constrained
to vanish because no dynamical quarks are present.

Working in Minkowski space, introduce the normalized singlet axial
current
\begin{align}
J_5^\mu(x)
=
\frac{1}{\sqrt{2N_f}}
\sum_{f=1}^{N_f}
\bar\psi_f(x)
\gamma^\mu\gamma_5
\psi_f(x),
\end{align}
which satisfies
\begin{align}
\partial_\mu J_5^\mu(x)
=
\frac{2i}{\sqrt{2N_f}}
\sum_{f=1}^{N_f}
m_f
\bar\psi_f
\gamma_5
\psi_f
+
\sqrt{2N_f}\,
q(x).
\label{Eq:anomaly_J50}
\end{align}
In the chiral limit this reduces to
\begin{align}
\partial_\mu J_5^\mu(x)
=
\sqrt{2N_f}\,
q(x),
\label{Eq:anomaly_chiral}
\end{align}
which is the anomalous singlet axial Ward identity underlying the
Witten--Veneziano relation.

The crucial large-$N_c$ observation is that, at fixed nonzero momentum,
quark-loop corrections to correlation functions of the gluonic
operator $q(x)$ are suppressed. The corresponding correlator in full
QCD therefore approaches that of pure Yang--Mills theory at leading
order in the $1/N_c$ expansion. The zero-momentum limit requires
greater care, however, because
\begin{align}
m_{\eta'}^2
=
{\cal O}(1/N_c),
\end{align}
so that the $\eta'$ pole produces a contribution that survives in the
topological susceptibility. Thus the limits $N_c\rightarrow\infty$
and $k\rightarrow0$ cannot simply be interchanged.

Defining the singlet decay constant by
\begin{align}
\braket{0|\hat J_5^\mu(0)|\eta'(k)}
=
if_0k^\mu,
\label{Eq:fetap_def}
\end{align}
one obtains
\begin{align}
\braket{0|
\partial_\mu\hat J_5^\mu(0)
|\eta'(k)}
=
f_0m_{\eta'}^2.
\label{Eq:div_coupling}
\end{align}
The $\eta'$ pole consequently contributes to the zero-momentum
correlator of the local topological charge density. At leading order
in the large-$N_c$ expansion, the full QCD susceptibility in the
chiral limit takes the form
\begin{align}
\chi_{\rm QCD}
=
\chi_{\rm YM}
-
\frac{f_0^2m_{\eta'}^2}{2N_f}
+
{\cal O}(1/N_c),
\label{Eq:chi_WV_cancel}
\end{align}
where the second term is the $\eta'$ pole contribution. The anomalous
Ward identity requires the full QCD susceptibility to vanish when one
or more quarks are exactly massless,
\begin{align}
\chi_{\rm QCD}=0.
\end{align}
Equation~(\ref{Eq:chi_WV_cancel}) therefore gives
\begin{align}
m_{\eta'}^2
=
\frac{2N_f}{f_0^2}\,
\chi_{\rm YM} [1+ {\cal O}(1/N_c)],
\end{align}
which is the Witten--Veneziano relation in the chiral limit. The
nonzero pure-glue susceptibility is thus canceled by the $\eta'$ pole
contribution in full massless QCD, while at the same time generating
a nonzero singlet pseudoscalar mass.

For the present discussion, the essential observation is that the
Witten--Veneziano relation requires the anomalous singlet axial Ward
identity, a nonvanishing zero-momentum correlation function of the
local operator $q(x)$ in pure Yang--Mills theory, and the large-$N_c$
expansion. The relevant quantity is the correlation function of the
local topological charge density rather than an \emph{a priori}
decomposition of the exact functional integral into integer-valued
topological sectors. The Witten--Veneziano relation therefore
demonstrates the physical significance of the local topological charge
density and the axial anomaly in generating the $\eta'$ mass, while
remaining logically independent of the additional assumptions leading
to a globally classified gauge-field configuration space.

\subsection{Leutwyler--Smilga relation}
\label{Sec:LS_relation}

The Leutwyler--Smilga relation connects the topological susceptibility
with the light quark masses in QCD. Like the Witten--Veneziano
relation, it follows from the anomalous Ward identity together with
the low-energy chiral effective theory. It therefore provides another
important example in which physically observable consequences of the
local topological charge density and its correlation functions may be
derived without, by themselves, requiring a global classification of
gauge-field configurations.

As discussed in Sec.~\ref{Sec:top_suscept}, the Euclidean generating
functional may be written as
\begin{align}
Z_E(\theta) = \textstyle \int \!
\mathcal D A_\mu
\mathcal D\bar\psi
\mathcal D\psi\,
e^{-S_E(\theta)},
\end{align}
where the parameter $\theta$ is introduced only as the source
parameter of the generating functional and is not interpreted as a
physical vacuum angle. The topological susceptibility is then obtained
from the vacuum energy density,
\begin{align}
\chi
&=
\left.
\frac{\partial^2\varepsilon_0(\theta)}
{\partial\theta^2}
\right|_{\theta=0},
\\
\varepsilon_0(\theta)
&\equiv
-
\lim_{V\rightarrow\infty}
\frac{\ln Z_E(\theta)}{V}.
\label{Eq:vareps_0_again}
\end{align}

At energies well below the chiral symmetry-breaking scale, the
generating functional is described by the chiral effective theory.
For the vacuum energy it is sufficient, at leading order, to consider
the constant vacuum alignment of the chiral field
$U\in SU(N_f)$. Using the anomalous axial change of variables discussed
in Sec.~\ref{Sec:theta_bar}, the source parameter $\theta$ may be 
represented equivalently in the quark mass matrix,
\begin{align}
M
\rightarrow
M_\theta
=
M e^{i\theta/N_f}.
\end{align}
The leading-order effective potential is then
\begin{align}
{\cal V}_{\rm eff}(U;\theta)
=
-
\Sigma\,
{\rm Re}\,
{\rm tr}
\left(
M e^{i\theta/N_f}U
\right),
\end{align}
where
\begin{align}
\Sigma
=
-\langle\bar\psi\psi\rangle_{\theta=0}
>0
\end{align}
is the chiral condensate.

For diagonal real positive quark masses, the vacuum alignment may be
written as
\begin{align}
U = \textstyle {\rm diag}
\left(
e^{i\alpha_1},
\ldots,
e^{i\alpha_{N_f}}
\right),
\qquad
\sum_{f=1}^{N_f}\alpha_f=0.
\end{align}
Defining
\begin{align}
\phi_f = (\theta/N_f)+\alpha_f,
\end{align}
one has
\begin{align}
\textstyle \sum_{f=1}^{N_f}\phi_f=\theta,
\end{align}
and the vacuum energy is therefore obtained by minimizing
\begin{align}
\varepsilon_0(\theta)
= \textstyle - \Sigma \max_{\{\phi_f\}} \left[
\sum_{f=1}^{N_f} m_f\cos\phi_f \right],
\qquad \sum_{f=1}^{N_f} \phi_f = \theta.
\end{align}
For notational simplicity we write $\theta$ rather than
$\bar\theta$ throughout this subsection, since the quark masses have
been chosen real and positive and only the dependence of the
generating functional on the source parameter is required.

For small $\theta$ one may expand
\begin{align}
\cos\phi_f = \textstyle 1 -\frac12 \phi_f^2 +
    {\cal O}(\phi_f^4),
\end{align}
so that
\begin{align}
\varepsilon_0(\theta) = \textstyle \varepsilon_0(0)
    +   \frac12 \Sigma
\min_{\{\phi_f\}}
\left[
\sum_{f=1}^{N_f}
m_f\phi_f^2
\right]
+
{\cal O}(m^2,\theta^4).
\end{align}
Minimization subject to the constraint
$\sum_f\phi_f=\theta$ gives
\begin{align}
\phi_f = \bar m\,\theta/m_f,
\qquad
\bar m^{-1} \equiv \textstyle  \sum_{f=1}^{N_f}
    (1/m_f),
\end{align}
and therefore
\begin{align}
\varepsilon_0(\theta) = \textstyle
    \varepsilon_0(0) + \frac12 \Sigma\bar m\, \theta^2
    + {\cal O}(m^2,\theta^4).
\end{align}
Taking the second derivative with respect to $\theta$ yields the
Leutwyler--Smilga relation,
\begin{align}
\chi
&=
\left.
\frac{\partial^2\varepsilon_0(\theta)}
{\partial\theta^2}
\right|_{\theta=0}
=
\Sigma\bar m
+
{\cal O}(m^2)
\nonumber\\
&=
\frac{\Sigma}
{\sum_{f=1}^{N_f}(1/m_f)}
+
{\cal O}(m^2).
\end{align}
For degenerate quark masses this reduces to
\begin{align}
\chi = m\Sigma/N_f + {\cal O}(m^2).
\end{align}
In particular, if any one quark mass tends to zero then
$\bar m\rightarrow0$ and hence $\chi\rightarrow0$, in agreement with
the anomalous chiral field-redefinition argument.

For the present discussion, the essential observation is that the
Leutwyler--Smilga relation follows from the generating functional, the
anomalous Ward identity, and the low-energy chiral effective theory.
The relevant quantity throughout the derivation is the dependence of
the vacuum energy on the source parameter $\theta$, or equivalently
the zero-momentum correlation function of the local topological charge
density. The relation therefore determines the dependence of the
topological susceptibility on the light quark masses without requiring
a nonzero physical value of $\theta$, or that the exact QCD functional
integral be fundamentally decomposed into configuration-by-configuration
sectors labeled by an integer-valued topological charge. Like the
Witten--Veneziano relation, it demonstrates the physical significance
of the local topological charge density while remaining logically
independent of the additional assumptions leading to a globally
classified gauge-field configuration space.

\subsection{Consequence of minimal QCD}
\label{Sec:What_established}

The preceding analysis has established the standard local
consequences of the topological charge density in QCD. These include
the construction of the generating functional $Z_E(\theta)$, the
topological susceptibility, the anomalous Ward identity, the
semiclassical role of instantons, the 't~Hooft effective interaction,
the Witten--Veneziano relation, and the Leutwyler--Smilga relation.
Collectively, these establish the known physical consequences of the
local topological charge density $q(x)$ and its correlation functions.

The essential observation is that none of these derivations has
required the assumption that the spacetime integral
$Q=\textstyle\int d^4x\,q(x)$ is an integer-valued topological
invariant, nor that the exact functional integral is fundamentally
decomposed into configuration-by-configuration sectors labeled by an
integer-valued topological charge. Rather, they rely on local gauge
invariance, correlation functions of local operators, the anomalous
Ward identity, semiclassical saddle-point methods where appropriate,
and, for the Witten--Veneziano relation, the large-$N_c$ expansion.

Throughout these derivations, the parameter $\theta$ has appeared
solely as an auxiliary source parameter of the generating functional
$Z_E(\theta)$; it has not been assumed to be a fixed physical vacuum
angle or coupling of the QCD action. Thus the source term proportional
to $\theta$ is introduced in the functional integral only for the
purpose of generating response functions. The required derivatives
with respect to the source are taken and the source is then set to
zero. A constant source $\theta$ generates the integrated, or
zero-momentum, correlation functions of $q(x)$; equivalently, one may
introduce a local source $\theta(x)$ to generate its general
correlation functions, with $\theta(x)$ likewise set to zero after
the required functional derivatives have been taken.

Within the framework of minimal QCD developed here, the additional
assumptions required to interpret
\begin{align}
Q
=
\textstyle
\int d^4x\,q(x)
\end{align}
as a globally defined integer-valued topological charge have not been
introduced. The physical action is therefore simply the standard QCD
action with real positive quark masses: it contains neither an
independent physical $\theta$ term nor an independent flavor-singlet
pseudoscalar quark mass term. Consequently, no physical
$CP$-violating parameter $\bar\theta$ has been introduced. The
auxiliary source $\theta$ appearing in the generating functional is
introduced only to generate correlation and response functions and,
after the required derivatives have been taken, is set to zero. It
does not constitute an additional parameter of the physical theory.

Within this framework, minimal QCD is therefore defined by the
standard QCD action with real positive quark masses, together with the
source-dependent generating functional used to define correlation and
response functions of the local operator $q(x)$. The anomalous axial
Ward identity establishes that, if either a physical $\theta$ term or
a flavor-singlet pseudoscalar mass term is introduced, the invariant
strong-sector $CP$-violating parameter is
\begin{align}
\bar\theta
=
\theta+\arg\det M.
\end{align}
This equivalence does not, however, require either contribution to be
present in the physical action. In minimal QCD neither is introduced,
and hence no physical $\bar\theta$ parameter is present. At this stage
the conventional strong $CP$ problem has therefore not arisen.

This naturally raises the following question: what additional
assumptions are required to pass from minimal QCD to extended QCD,
and, in particular, before one may consistently interpret $Q$ as an
integer-valued global topological invariant, promote the auxiliary
source $\theta$ to a periodic physical vacuum angle, and construct
the conventional $\theta$-vacuum framework? We now turn to that
question.

\section{Additional Assumptions Leading to Global Topological Classification}
\label{Sec:Global_Classification}

The preceding section established that the known local consequences
of the topological charge density in QCD follow without assuming
that the exact quantum functional integral is fundamentally
restricted to smooth finite-action gauge fields admitting a
global topological classification. The purpose of this section is
not to question the mathematical validity of the conventional
construction, but to identify precisely the point at which the
\emph{additional assumptions} are introduced. We now examine these
assumptions and show how they lead to the conventional global
topological formulation of QCD together with its associated physical
consequences.

\subsection{Smooth gauge fields and global topological classification}

The central question is not whether sufficiently smooth
finite-action gauge fields, with appropriate global boundary
conditions, admit an integer-valued topological classification. They
do. Rather, the question is whether the exact QCD functional integral
is fundamentally defined only over configurations possessing this
smooth global structure, or whether imposing such structure
constitutes an additional assumption supplementing the local
formulation of QCD.

As mentioned above, in the functional integral typical quantum fields
are distributions rather than smooth classical fields
\citep{Glimm:1987ylb,Simon:2015spl,simon2005functional,
colella1973sample}. Such configurations are not, in general,
differentiable in the classical sense and need not satisfy the
regularity conditions required for the usual global topological
classification of gauge fields. Consequently, the existence of
smooth finite-action gauge fields with integer-valued topological
charge does not by itself establish that the exact quantum functional
integral is restricted to only include such configurations.

The same issue appears in lattice QCD. The continuum limit does not
make typical ultraviolet gauge-field configurations pointwise smooth.
To associate a robust integer-valued topological charge with lattice
configurations, additional constructions are ordinarily employed. One
possibility is to smooth the gauge field using, for example, gradient
flow. Another is to employ an index-theoretic definition based on an
appropriate lattice Dirac operator
\citep{Neuberger:1998,Hasenfratz:1998}. For configurations on which
such an operator is well-defined, its index provides an
integer-valued lattice topological charge without requiring the
original link variables themselves to be smooth classical fields.

An index-theoretic definition may therefore assign an integer-valued
charge $Q$ to a rough lattice configuration, but this does not by
itself establish the pointwise continuum identity
\begin{align}
Q
=
\frac{g^2}{16\pi^2}
\textstyle
\int d^4x\,
\mathrm{tr}
\!(F_{\mu\nu}\tilde F^{\mu\nu})
=
\textstyle
\int d^4x\,q(x)
\end{align}
for an arbitrary rough continuum configuration. Recovering this
continuum expression requires an appropriate continuum limit together
with the regularity and renormalization properties needed to define
the corresponding local operator and its spacetime integral.

Thus lattice QCD demonstrates that well-defined and physically useful
integer-valued notions of topology can be constructed nonperturbatively,
but it does not by itself imply that the unsmoothed continuum
functional integral is literally an integral over smooth classical
gauge fields carrying a configuration-by-configuration integer-valued
charge.

The conventional global classification may be formulated in several
closely related settings. On Euclidean $\mathbb R^4$, sufficiently
smooth finite-action gauge fields approach a pure gauge at spacetime
infinity. After compactifying the boundary, the asymptotic gauge
transformation defines a smooth map
\begin{align}
U:S^3\rightarrow SU(3),
\end{align}
whose homotopy class is classified by
\begin{align}
\pi_3(SU(3))=\mathbb Z.
\end{align}
Equivalently in spirit, but mathematically as a distinct global
construction, one may formulate the theory on a compact four-manifold
and classify smooth principal $SU(3)$ bundles by their characteristic
classes. On $S^4$, for example, the relevant second Chern number is
integer valued; analogous bundle constructions may be used on $T^4$,
with the appropriate global boundary conditions and bundle data.

Under these smoothness and global assumptions one obtains
\begin{align}
Q
=
\textstyle
\int d^4x\,q(x)
\in\mathbb Z,
\end{align}
with $Q$ identified with the corresponding characteristic number.
The space of admissible smooth gauge configurations may then be
organized into components or bundle classes labeled by this integer,
and the functional integral correspondingly written as a sum over
topological sectors. Within this framework, the Atiyah--Singer index
theorem relates the index of the Dirac operator to the same
topological charge.

Once this additional global structure is adopted, one may weight the
integer sectors by
\begin{align}
e^{i\theta Q},
\end{align}
so that $\theta$ acquires its conventional interpretation as a
periodic angular parameter, $\theta\simeq\theta+2\pi$, and the
$\theta$-vacuum construction follows. The local coupling
$\theta q(x)$ can of course be written without first assuming that
$Q$ is integer valued; what depends upon the global topological
classification is its interpretation as an angular vacuum parameter
weighting integer-valued sectors and the associated conventional
$\theta$-vacuum framework.

\subsection{Topological sectors and the $\theta$ parameter}

Once the space of smooth gauge-field configurations has been
decomposed into integer-valued topological sectors, the functional
integral may be written as
\begin{align}
\mathcal{C}
=
\textstyle
\bigcup_{Q\in\mathbb Z}
\mathcal{C}_Q ,
\end{align}
where $\mathcal{C}_Q$ denotes the set of smooth gauge-field
configurations with topological charge $Q$. The Euclidean generating
functional therefore takes the form
\begin{align}
Z_E
=
\textstyle
\sum_{Q\in\mathbb Z}
\int_{\mathcal C_Q}
\mathcal D A_\mu\,
\mathcal D\bar\psi\,
\mathcal D\psi\,
e^{-S_E[A,\psi,\bar\psi]} .
\end{align}

Having decomposed the functional integral into disconnected
topological sectors, one may ask whether the different sectors should
enter with equal weight or whether they may carry relative phase
factors. More generally, assign a weight $w(Q)$ to each sector,
obtaining
\begin{align}
\tilde Z_E
=
\textstyle
\sum_{Q\in\mathbb Z}
w(Q)
\int_{\mathcal C_Q}
\mathcal D A_\mu\,
\mathcal D\bar\psi\,
\mathcal D\psi\,
e^{-S_E[A,\psi,\bar\psi]} .
\end{align}

The allowed form of $w(Q)$ is constrained by locality. For two
well-separated spacetime regions carrying topological charges
$Q_1$ and $Q_2$, additivity requires
\begin{align}
Q
=
Q_1+Q_2 ,
\end{align}
while locality requires the corresponding weights to factorize,
\begin{align}
w(Q_1+Q_2)
=
w(Q_1)\,
w(Q_2).
\end{align}
If the relative weights are required to be phases, the allowed
weights are therefore the one-dimensional unitary representations
of the additive group $\mathbb Z$,
\begin{align}
w(Q)
=
e^{i\theta Q},
\qquad
\theta\simeq\theta+2\pi,
\end{align}
up to an overall normalization. Consequently, the functional integral
takes the form
\begin{align}
\tilde Z_E(\theta)
=
\textstyle
\sum_{Q\in\mathbb Z}
e^{i\theta Q}
\int_{\mathcal C_Q}
\mathcal D A_\mu\,
\mathcal D\bar\psi\,
\mathcal D\psi\,
e^{-S_E[A,\psi,\bar\psi]}
\equiv
\textstyle
\sum_{Q\in\mathbb Z}
e^{i\theta Q}
Z_E^Q .
\label{Eq:QCD_tildeZ}
\end{align}

This construction also makes several properties of
$\tilde Z_E(\theta)$ immediate. The functional measure
$\mathcal D A_\mu$ integrates over every gauge-field configuration
together with its $CP$ conjugate and is therefore $CP$ invariant at
$\theta=0$. Since the topological charge density is $CP$ odd, a
$CP$ transformation sends
\begin{align}
Q\rightarrow -Q.
\end{align}
For real positive quark masses and in the absence of any other
$CP$-violating interaction, the functional measure and action at
$\theta=0$ are $CP$ invariant.
It follows that the sector partition functions satisfy
\begin{align}
Z_E^Q=Z_E^{-Q},
\end{align}
and hence
\begin{align}
\tilde Z_E(\theta)
&=
\textstyle
\sum_{Q\in\mathbb Z}
e^{i\theta Q}Z_E^Q
=
\textstyle
\sum_{Q\in\mathbb Z}
e^{-i\theta Q}Z_E^Q
=
\tilde Z_E(-\theta),
\end{align}
so that $\tilde Z_E(\theta)$ is an even function of $\theta$.

Furthermore, since $Q$ is integer valued, the functional integral is
periodic under
\begin{align}
\theta\rightarrow\theta+2\pi,
\end{align}
with
\begin{align}
\tilde Z_E(\theta+2\pi)
=
\tilde Z_E(\theta).
\end{align}
Thus $\theta$ becomes an angular variable with period $2\pi$. It is
this periodicity, together with the decomposition of the functional
integral into integer-valued topological sectors, that leads to the
conventional $\theta$-vacuum formulation of QCD.

\subsection{The local $\theta$ term}

Having decomposed the functional integral into integer-valued
topological sectors, one may rewrite the relative sector weights in a
local form. For sufficiently smooth gauge fields, the topological
charge may be expressed as
\begin{align}
Q
=
\textstyle
\int d^4x\,q(x),
\end{align}
where
\begin{align}
q(x)
=
\frac{g^2}{16\pi^2}
\,
\mathrm{tr}
\!\left(
F_{\mu\nu}\tilde F^{\mu\nu}
\right)
\end{align}
is the local topological charge density introduced in
Sec.~\ref{Sec:local_QCD}. The sector weight may therefore be written
as
\begin{align}
e^{i\theta Q}
=
\exp
\!\left[
i\theta
\textstyle
\int d^4x\,q(x)
\right].
\end{align}
The sector-weighted generating functional consequently becomes
\begin{align}
\tilde Z_E(\theta)
=
\textstyle
\sum_Q
\int_{\mathcal C_Q}
\mathcal D A_\mu\,
\mathcal D\bar\psi\,
\mathcal D\psi\,
\exp
\!\left[
-
S_E
+
i\theta
\int d^4x\,q(x)
\right].
\end{align}

For sufficiently smooth gauge fields this expression is equivalent to
introducing the local Euclidean Lagrangian density
\begin{align}
\mathcal L_E(\theta)
=
\mathcal L_E
-
i\theta q(x),
\end{align}
or, equivalently, the Euclidean action
\begin{align}
S_E(\theta)
=
S_E
-
i\theta
\textstyle
\int d^4x\,q(x).
\end{align}

Thus, once the additional assumptions leading to an
integer-valued topological classification have been adopted, the
sector-weighted formulation and the familiar local $\theta$ term are
mathematically equivalent descriptions of the same theory. A local
coupling to $q(x)$ can of course be written without first introducing
this global construction. What follows specifically from the global
classification is the interpretation of this coupling as the local
representation of the relative phase $e^{i\theta Q}$ assigned to
integer-valued topological sectors, together with the resulting
$2\pi$ periodicity and vacuum-angle interpretation of $\theta$.

\subsection{The $\theta$ vacuum}

The Euclidean topological sectors discussed above classify
four-dimensional gauge-field histories by their integer-valued
topological charge $Q$. In the conventional Hamiltonian construction,
a related but distinct integer arises on a spatial hypersurface. Pure
gauge vacuum configurations are labeled by an integer winding number
$n$, equivalently by the Chern--Simons number modulo large gauge
transformations. A Euclidean configuration interpolating between
vacuum sectors $n_i$ and $n_f$ carries topological charge
\begin{align}
Q
=
n_f-n_i .
\end{align}
Thus the integer $Q$ labeling a four-dimensional Euclidean history
should not be identified directly with the integer $n$ labeling a
vacuum configuration on a spatial hypersurface.

The conventional $\theta$ vacuum is then constructed as
\begin{align}
\ket{\Omega_\theta}
=
\textstyle
\sum_{n\in\mathbb Z}
e^{in\theta}
\ket{n},
\end{align}
up to the convention chosen for the sign of $\theta$. A detailed
discussion of this Hamiltonian construction, including the role of
large gauge transformations and the relation between winding number,
Chern--Simons number, and four-dimensional topological charge, is
given in Ref.~\cite{Williams:2026cec}.

For the present discussion, the essential point is that a history
connecting sectors $n_i$ and $n_f$ has $Q=n_f-n_i$. The relative
phase between its initial and final vacuum components therefore
reproduces the factor $e^{i\theta Q}$ appearing in the Euclidean
sector-weighted functional integral of
Eq.~(\ref{Eq:QCD_tildeZ}). The Hamiltonian $\theta$-vacuum
construction and the sector-weighted Euclidean functional integral
are therefore the conventional complementary realizations of the
same assumed global topological structure.

\subsection{Emergence of the conventional strong $CP$ problem}

Sections~\ref{Sec:local_QCD}--\ref{Sec:What_established}
established that the known local consequences of the topological
charge density follow from minimal QCD. In particular, the local
operator $q(x)$, the generating functional $Z_E(\theta)$ viewed as a
functional of an auxiliary source parameter, the topological
susceptibility, the anomalous Ward identity, the semiclassical role of
instantons, the 't~Hooft effective interaction, the
Witten--Veneziano relation, the Leutwyler--Smilga relation, and the
corresponding lattice QCD results were all obtained without assuming
that the exact functional integral is fundamentally restricted to
smooth gauge fields admitting a global integer-valued topological
classification. In these constructions the required derivatives with
respect to the auxiliary source are taken and the source is then set
to zero; it is not an additional parameter of the physical theory.

The present section has examined the additional assumptions that lead
to the conventional global topological formulation of QCD.
Restricting the functional integral to sufficiently smooth
finite-action gauge fields with the appropriate global structure
permits the introduction of an integer-valued topological charge,
\begin{align}
Q
=
\textstyle
\int d^4x\,q(x)
\in\mathbb Z,
\end{align}
together with a decomposition of the functional integral into
topological sectors, the relative sector weighting
$e^{i\theta Q}$, its equivalent representation by a local $\theta$
term in the action, and the corresponding $\theta$ vacuum. The
integer-valued nature of $Q$ makes $\theta$ a periodic angular
parameter, with $\theta\simeq\theta+2\pi$.

Within this enlarged framework, $\theta$ is no longer merely the
auxiliary source parameter of the generating functional
$Z_E(\theta)$, which is set to zero after differentiation. Rather, it
acquires the interpretation of a physical periodic vacuum angle.
Since the operator $q(x)$ is odd under $CP$, a $CP$ transformation
maps
\begin{align}
\theta\rightarrow-\theta.
\end{align}
Together with the $2\pi$ periodicity, this implies that the action is
$CP$ invariant at $\theta=0$ and $\theta=\pi$ modulo $2\pi$,
although at $\theta=\pi$ the vacuum may spontaneously break $CP$.
For a generic value of $\theta$, the $\theta$ term explicitly
violates $CP$.

The most general quark mass matrix may contain a flavor-singlet
pseudoscalar term. The anomalous axial change of fermion variables
relates the gluonic $\theta$ term and the corresponding pseudoscalar
quark mass term, leaving the invariant combination
\begin{align}
\bar\theta
=
\theta+\arg\det M.
\end{align}
Thus, once an independent physical $\bar\theta$ parameter is admitted,
the observable strong-sector $CP$ violation is governed by
$\bar\theta$, while its representation in terms of $\theta$ and the
pseudoscalar mass term is convention dependent.
It is precisely within extended QCD that the conventional strong
$CP$ problem arises: strong interactions are experimentally
constrained to conserve $CP$ to very high precision, requiring the
physical parameter $\bar\theta$ to be extraordinarily small.

The logical distinction between Sections II and III is therefore
straightforward. Section II established the consequences of minimal
QCD, based on local gauge invariance, causal locality, and the
standard functional integral. In minimal QCD neither an independent
physical $\theta$ term nor an independent flavor-singlet pseudoscalar
mass term is introduced, and hence no physical $\bar\theta$ parameter
is present. Extended QCD admits such an independent physical
parameter. Section III has examined, in particular, the additional
global assumptions concerning the smoothness and topological
classification of gauge fields that lead to the conventional
$\theta$-vacuum construction and the interpretation of $\theta$ as a
physical vacuum angle. Once extended QCD is adopted, the subsequent
study of $\bar\theta$, its phenomenological consequences, and possible
mechanisms accounting for its smallness occurs entirely within that
framework. The extensive literature devoted to these questions
therefore addresses issues that arise once an independent physical
$\bar\theta$ parameter has been admitted; for example, see
Refs.~[6--26]. The present work does not question those analyses; it
asks only whether established physical observables require such an
independent physical parameter and hence require going beyond minimal
QCD.

\section{Conclusions}
\label{Sec:Conclusions}

The conventional strong $CP$ problem arises within what we have
called here \emph{extended QCD}, in which an independent physical 
flavor-singlet $CP$-violating parameter $\bar\theta$ is admitted.
In the conventional global topological construction, sufficiently
smooth finite-action gauge fields with appropriate global boundary
conditions possess the integer-valued topological invariant
\begin{align}
Q
=
\textstyle
\int d^4x\,q(x).
\end{align}
The functional integral may then be decomposed into sectors labeled
by $Q\in\mathbb Z$, with relative weights $e^{i\theta Q}$. This makes
$\theta$ a periodic angular variable, permits its interpretation as a
physical vacuum angle, and leads to the conventional $\theta$-vacuum
construction. More generally, the same physical strong-sector
$CP$-violating parameter may be represented by a gluonic $\theta$
term, a flavor-singlet pseudoscalar phase of the quark mass matrix,
or a combination of the two. The anomalous axial Ward identity
identifies the invariant combination as
\begin{align}
\bar\theta
=
\theta+\arg\det M.
\end{align}
The experimental absence of strong $CP$ violation then requires
$\bar\theta$ to be extraordinarily small, giving the conventional
strong $CP$ problem.

The present work has asked a logically prior question: which
established consequences of QCD actually require extending QCD in
this way? To address this question we have distinguished two
formulations. We defined here for convenience \emph{minimal QCD} as the
standard QCD theory with real positive quark masses, local gauge
invariance, causal locality, and the corresponding functional
integral, without introducing an independent physical flavor-singlet
$CP$-violating parameter $\bar\theta$ or assuming a global
integer-valued classification of gauge-field configurations. We
defined \emph{extended QCD} as formulations in which an independent
physical flavor-singlet $CP$-violating parameter $\bar\theta$ is
introduced. Thus the physical action of minimal QCD contains neither
an independent physical $\theta$ term nor an independent
flavor-singlet pseudoscalar quark mass term, whereas extended QCD
admits the corresponding invariant physical parameter
$\bar\theta=\theta+\arg\det M$.

This definition does not prevent the use of $\theta$ as an auxiliary
source. A source term proportional to $\theta q(x)$ may be introduced
in the generating functional in order to define correlation and
response functions of the local topological charge density. The
required derivatives with respect to the source are taken and the
source is then set to zero. A constant source generates integrated,
or zero-momentum, correlation functions, including the topological
susceptibility, while a local source $\theta(x)$ may be used to
generate general correlation functions of $q(x)$. Such an auxiliary
source is not an additional parameter of the physical theory.

Within minimal QCD, the established consequences associated with the
local topological charge density remain intact. These include the
topological susceptibility, the anomalous singlet axial Ward identity,
the semiclassical role of instantons, the 't~Hooft effective
interaction, the Witten--Veneziano relation, the
Leutwyler--Smilga relation, and their nonperturbative realization in
lattice QCD. These results demonstrate the physical significance of
$q(x)$ and its correlation functions, but none requires the exact
functional integral to be fundamentally decomposed into
configuration-by-configuration sectors carrying an integer-valued
topological charge.

This distinction is particularly important because typical quantum
fields in the functional integral are not smooth classical
configurations. Smooth finite-action fields provide an essential
semiclassical description and possess a well-defined global
topological classification under the appropriate boundary
conditions, but this does not by itself establish that the exact
functional integral is fundamentally an integral only over
configurations possessing that classical global structure. Likewise,
lattice QCD provides powerful nonperturbative definitions of
topological observables, through smoothing procedures such as
gradient flow or index-theoretic constructions, without establishing
that an arbitrary unsmoothed continuum configuration possesses a
configuration-by-configuration integer value of
$\int d^4x\,q(x)$ in the classical sense.

If the additional smoothness and global assumptions of the
conventional construction are adopted within extended QCD, the
familiar global topological formulation follows. One obtains an
integer-valued $Q$, a decomposition into topological sectors, relative
weights $e^{i\theta Q}$, $2\pi$ periodicity in $\theta$, and the
conventional $\theta$ vacuum. The local term $\theta q(x)$ can of
course be written without first imposing this global classification.
What the global construction supplies is its interpretation as the
local representation of the relative phases assigned to
integer-valued topological sectors and, consequently, the
interpretation of $\theta$ as a periodic physical vacuum angle.

The local axial anomaly provides a separate and exact statement. If
an independent flavor-singlet pseudoscalar phase is admitted in the
quark mass matrix, an anomalous axial change of fermion variables
transfers the phase between the quark mass matrix and a gluonic
$\theta$ term, leaving the invariant combination
\begin{align}
\bar\theta
=
\theta+\arg\det M.
\end{align}
The present analysis does not question this equivalence. Nor does it
claim that introducing an independent physical flavor-singlet
$CP$-violating parameter is inconsistent with the local symmetries of
QCD. Rather, the present work distinguishes the statement that such a
parameter \emph{may} consistently be introduced from the stronger
statement that established QCD phenomenology \emph{requires} it to
be present.

The principal conclusion of the present work is therefore direct.
Minimal QCD, as defined here, contains no independent physical
$\theta$ term, no independent flavor-singlet pseudoscalar mass term,
and hence no physical $\bar\theta$ parameter. Nevertheless, the known
physical consequences associated with the local topological charge
density and the axial anomaly are recovered. We are unaware of any
established physical observable that requires extending minimal QCD
by introducing an independent physical flavor-singlet $CP$-violating
parameter $\bar\theta$. Nor are we aware of any such observable that
requires the additional global assumptions leading to an
integer-valued topological classification and the conventional
$\theta$-vacuum interpretation. The conventional strong $CP$ problem
arises within extended QCD once an independent physical $\bar\theta$
is admitted.

This conclusion does not question the mathematical consistency or
the extensive phenomenology developed within extended QCD. Once a
physical $\bar\theta$ is admitted, its experimental smallness presents
the familiar fine-tuning problem known as the strong $CP$ problem.
The subsequent study of neutron electric dipole moments,
Peccei--Quinn symmetry, axions, Nelson--Barr models, lattice topology,
and related nonperturbative phenomena then proceeds exactly as in the
conventional literature. The present work addresses the logically
prior question of whether any established physical observable
requires the introduction of the independent physical
$\bar\theta$ parameter upon which this problem is predicated.
The presumption that the strong $CP$ problem is
\emph{unavoidable} therefore follows only if a consistent definition
of QCD \emph{requires} an independent physical $\bar\theta$
parameter. At this time there does not appear to be any physical
observable or foundational principle that provides such a
requirement.

Future theoretical developments or experimental discoveries could
provide such a requirement. If they do, extended QCD and the
extensive body of work developed within it apply immediately. In the
absence of such a requirement, however, the strong $CP$ problem
should not be regarded as an unavoidable consequence of the local
structure of QCD itself. It arises upon extending minimal QCD by
introducing an independent physical flavor-singlet $CP$-violating
parameter $\bar\theta$. This parameter may be represented by a
gluonic $\theta$ term, a flavor-singlet pseudoscalar phase of the
quark mass matrix, or a combination of the two, with the anomalous
axial Ward identity relating these representations through the
invariant combination
$\bar\theta=\theta+\arg\det M$.

\begin{acknowledgments}
The author is grateful to Björn Garbrecht, Martin L\"uscher, 
Andreas Ringwald, Francesco Sannino and Gerrit Schierholz for 
valuable discussions that helped clarify several aspects of the
presentation. The conclusions of the paper are, of course, the
responsibility of the author alone. AGW is funded by the ARC 
Centre of Excellence for Dark Matter Particle Physics (CDMPP)
CE200100008 and further supported by the Centre for the 
Subatomic Structure of Matter (CSSM). 
\end{acknowledgments}

\appendix


\bibliography{theta_refs}

\end{document}